# When the macroscopic game is a quantum game?

A.A.Grib, G.N.Parfionov

Abstract

An example of the macroscopic game of two partners consisting of two classical games played simultaneously with special dependence of strategies is considered. The average profit of each partner is equal to the average profit obtained in the quantum game with two noncommuting operators for the spin one half system with strategies defined by the wave function. Nash equilibria in the macroscopic game coincide with Nash equilibria of the quantum game.

Introduction

In our previous works [1,2] we discussed the possibility of macroscopic games described not by the standard Kolmogorovian probability measure but by the wave function as some vector in Hilbert space. Only examples for finite dimensional spin one half and spin one were considered.

However the problem of interpretation of such examples was still open. Here we consider some simple enough cases of macroscopic games as quantum games. The quantum game, either "microscopic" (i.e. using quantum microparticles as it is supposed in the majority of works on quantum games) or macroscopic (without any microparticles) deals with some quantum logical nondistributive orthocomplemented lattices. For such lattices due to nondistributivity property one cannot define one Kolmogorovian probability measure for all elements but can define the so called "quantum measure", i.e. the wave function as some vector in finite dimensional Hilbert space. Definite wave function leads to definite weights (some nonnegative numbers) for each atom of the lattice. For distributive sublattices of the initial lattice these weights have the meaning of Kolmogorovian probability measures.

So due to Born interpretation of quantum mechanics the wave function defines different probability measures for different complementary noncommuting observables and weights are statistically interpreted as frequencies of getting positive answers for values of observables. This means that for these finite dimensional cases instead of the wave function one can speak about the set of weights or frequencies and v.v.

Search for macroscopic quantum games means looking for situations when such set of weights can occur. Analyzing the rule of calculation of the average profit in any quantum game one can see [1,2] that it is calculated as the expectation value of the sum of noncommuting operators for the wave function as the tensor product of wave functions for each participant of the game. It occurs that it is the sum of the products of normalized weights on the elements of the payoff matrix. For the case of two noncommuting observables the sum of weights is equal to 2, for three observables it is 3 etc.

But then the simple idea arises to consider one quantum game as many classical games, described by Boolean sublattices of the orthocomplemented lattice with special prescription for probability distributions. In quantum case different Boolean sublattices correspond in spin cases to different noncommuting observables where the difference is due to taking projections of different spin observables. This difference is described by the angle. So weights (or frequencies) are parametrized by this angle and are not independent. In [1,2] some simplified version with real two dimensional case was used. The connection of weights and the wave function and observables for spin one half case was considered in [3].

For the case of the quadrangle spin one half game [1,2] when two noncommuting observables, for example the spin projection on z-axis and the projection on some other direction parametrized by the angle $\theta$ are measured for some given wave function the frequencies for getting this or that answer for one projection can be taken as arbitrary equal to some $\sin^2\alpha$ and $\cos^2\alpha$. But then the frequencies for the other projections are not arbitrary! They must be $\sin^2(\alpha - \theta)$ and $\cos^2(\alpha - \theta)$. The game is organized in such a way that the angle $\theta$ is fixed and only $\alpha$ is variable.

So if some macroscopical player Alice is playing two games at once, using for her strategies probabilities different for different games where the difference is described just by our quantum rule, then this will be our quantum game. The profit is calculated as the sum of profits in two games and the average profit will be given just by the quantum expectation value as it was in [1,2]. Nash equilibria for such two games considered as one game can be found by the rule in [1,2].

In macroscopic situations quantum games occur due to special form of dependence of strategies in different classical games. This dependence can be due to some asymmetry in acts of the player simultaneously playing different classical games. For example he (she) cannot have the same frequency for acts done by his right or left hand etc. For the quantum game when three noncommuting spin observables are measured this dependence can be manifested in Heisenberg uncertainty relations for spin written in the form of some relations for frequencies in three classical games.

So "quantum casino" can be organized if its owner just asks the players to follow the quantum rule for the frequencies. This rule can be due to use of some "hardware" leading to asymmetry in acts of players, playing many games at once.

More generally one can speak about some specific "quantum correlation" occurring for what we call macroscopic quantum games. The player plays two classical games at once and the strategies used by him in these games are such that the probabilities in one and the other game at satisfy some special condition (a nonlinear one) taken from the existence of the wave function and noncommuting observables. The average profit got in two games is equal to that calculated by the rules of quantum mechanics. One can forget about quantum physics at all and speak about he situation when there is exchange of information for one player from one game to the other manifested in dependence of strategies in two games expressed by the condition for probabilities.

It is interesting if such quantum game situations can occur in Nature in biological evolution etc?

## 1. The model of the game

Here we give formulas for the quantum spin one half game with two noncommuting observables and corresponding formulas for two classical games leading to the same Nash equilibria due to special dependence of probabilities for strategies in these two games. The average profit of Alice playing with Bob is calculated in [1,2] as the expectation value (1) of the payoff operator H being the sum of noncommuting projection operators $A_i$, $B_k$:

$$H = c_3 A_1 \otimes B_3 + c_1 A_3 \otimes B_1 + c_4 A_2 \otimes B_4 + c_2 A_4 \otimes B_2$$

$$\langle H \rangle = \langle \varphi | \otimes \langle \psi | H | \psi \rangle \otimes | \varphi \rangle = c_3 p_1 q_3 + c_1 p_3 q_1 + c_4 p_2 q_4 + c_2 p_4 q_2 \qquad (1)$$

where $\varphi, \psi$ – normalized vectors in real two-dimensional space, $p_i = \langle \varphi | A_i | \varphi \rangle$, $q_i = \langle \psi | B_i | \psi \rangle$. Because one has:

$$A_1 + A_3 = I, \quad A_2 + A_4 = I, \quad B_1 + B_3 = I, \quad B_2 + B_4 = I$$

then

$$p_1 + p_3 = 1, \quad p_2 + p_4 = 1, \quad q_1 + q_3 = 1, \quad q_2 + q_4 = 1 \qquad (2)$$

The wave functions are taken as unit vectors:

$$\varphi = (\cos\alpha, \sin\alpha), \qquad \psi = (\cos\beta, \sin\beta)$$

Then one obtains for probabilities:

$$p_1 = \cos^2\alpha, \quad p_2 = \cos^2(\alpha - \theta), \quad q_1 = \cos^2\beta, \quad q_2 = \cos^2(\beta - \tau) \qquad (3)$$

The main idea of this paper is that the same expression for the average profit calculated for the quantum game can be obtained if one is considering two games played by each partner simultaneously.

Let Alice is playing the games on two desks: one called "even", the other one "odd". The same is for Bob. The average profits for Alice in each of the parallel games are

$$h_{odd} = c_3 p_1 \cdot q_3 + c_1 p_3 \cdot q_1, \qquad h_{even} = c_4 p_2 \cdot q_4 + c_2 p_4 \cdot q_2$$

So for the average profit in two games one obtains

$$h = c_3 p_1 \cdot q_3 + c_1 p_3 \cdot q_1 + c_4 p_2 \cdot q_4 + c_2 p_4 \cdot q_2$$

The important feature of these classical games making them different from well known situations is the existence of "quantum cooperation" given by formula (3) with fixed $\theta, \tau$. This "cooperation" can be written in more symmetric form as some equation for $p_1, p_2$. To do that introduce:

$$x_1 = 1 + p_1 + p_2, \qquad x_2 = -p_1 + p_2$$

Then after some trigonometric operations one obtains from (3)

$$\frac{x_1^2}{\cos^2\theta} + \frac{x_2^2}{\sin^2\theta} = 1 \qquad (4)$$

i.e. equation of the ellipse with axes defined by $\cos\theta$, $\sin\theta$.

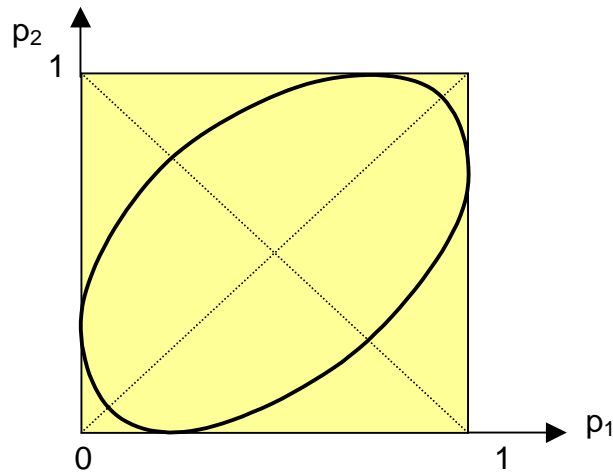

Figure.1. *The strategies of Alice in two games correspond to points of the ellipse*

In parametric form the eq. (4) can be written as:

$$\frac{x_1}{\cos\theta} = \cos(2\alpha - \theta), \qquad \frac{x_2}{\sin\theta} = \sin(2\alpha - \theta)$$

Nash equilibria can be found as:

$$\langle H \rangle = c_3 p_1 q_3 + c_1 p_3 q_1 + c_4 p_2 q_4 + c_2 p_4 q_2 \quad \rightarrow \quad \max_p \min_q \qquad (5)$$

with conditions

$$p_1 + p_3 = 1, \qquad p_2 + p_4 = 1, \qquad (6)$$

$$p_1^2 + p_2^2 - 2p_1 p_2 \cos 2\theta - p_1[\sin^2 2\theta - 2\sin^2\theta \cos 2\theta] - 2p_2 \sin^2\theta + \sin^4\theta = 0, \qquad (7)$$

$$q_1 + q_3 = 1, \qquad q_2 + q_4 = 1, \qquad (8)$$

$$q_1^2 + q_2^2 - 2q_1 q_2 \cos 2\tau - q_1[\sin^2 2\tau - 2\sin^2\tau \cos 2\tau] - 2q_2 \sin^2\tau + \sin^4\tau = 0 \qquad (9)$$

Here the conditions (6–9) express "quantum cooperation" for strategies. This is not the same as classical correlation and is some new feature of our game.

Really if one considers two games as one antagonistic classical game the possible strategies can be considered as "1-2", "3-2", "1-4".

Table 1. *Payoff-matrix of two-games as one antagonistic classical game*

|  | 1-2$_B$ | 1-4$_B$ | 3-2$_B$ | 3-4$_B$ |
|---|---|---|---|---|
| 1-2$_A$ | 0 | $c_2$ | $c_1$ | $c_1+c_2$ |
| 1-4$_A$ | $c_4$ | 0 | $c_1+c_4$ | $c_1$ |
| 3-2$_A$ | $c_3$ | $c_3+c_2$ | 0 | $c_2$ |
| 3-4$_A$ | $c_3+c_4$ | $c_3$ | $c_4$ | 0 |

Same is for Bob. Then introducing mixed strategies of Alice and Bob in this classical matrix game as $p_{ik}, q_{ik}$ one has:

$$p_1 = p_{12} + p_{14}, \quad p_3 = p_{32} + p_{34}, \quad p_2 = p_{12} + p_{32}, \quad p_4 = p_{14} + p_{34} \qquad (10)$$

It is evident that $p_1 + p_3 = 1$, $p_2 + p_4 = 1$. Acts of Alice in two games can be independent, i.e.:

$$p_{12} = p_1 \cdot p_2, \quad p_{32} = p_3 \cdot p_2, \quad p_{14} = p_1 \cdot p_4, \quad p_{34} = p_3 \cdot p_4 \qquad (11)$$

But eq. (7, 9) can be still valid, showing the difference of "quantum correlation" from the classical one expressed by breaking conditions (11).

## 3. The complex case

In the general complex two-dimensional case for spin one half system with two noncommuting observables one has:

$$A_1 = \begin{Vmatrix} 1 & 0 \\ 0 & 0 \end{Vmatrix} \qquad A_2 = \begin{Vmatrix} \cos^2\theta & \sin\theta\cos\theta\, e^{i\lambda} \\ \sin\theta\cos\theta\, e^{-i\lambda} & \sin^2\theta \end{Vmatrix}$$

The wave function is some vector $\varphi = e^{i\omega}(\cos\alpha, \sin\alpha)$. Real case considered in section 2 means $\lambda = 0$. In general case one has:

$$p_1 = \langle\varphi|A_1|\varphi\rangle = \cos^2\alpha,$$

$$p_2 = \langle\varphi|A_3|\varphi\rangle = \cos^2\alpha\cos^2\theta + \sin^2\alpha\sin^2\theta + 2\cos\alpha\cos\theta\sin\alpha\sin\theta\cdot\cos\lambda$$

It can be seen that $p_1, p_2$ are again located on the ellipse with orientation and localization dependent on $\lambda$.

$$p_1^2[\cos^2 2\theta + \sin^2 2\theta \cos^2\lambda] + p_2^2 - 2p_1p_2\cos 2\theta - \\ - p_1[\sin^2 2\theta - 2\sin^2\theta\cos 2\theta] - 2p_2\sin^2\theta + \sin^4\theta = 0 \qquad (12)$$

Same equations are valid for the other partner with some other angle $\tau$ instead of $\theta$ and $\mu$ instead of $\lambda$.

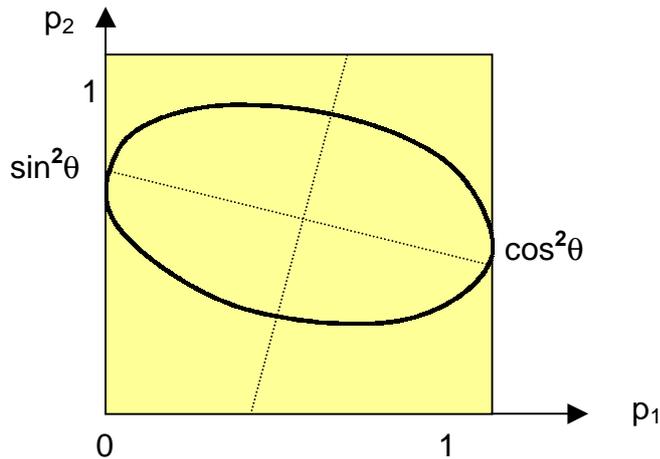

Figure 2. *The ellipse is located inside the unit quadrangle and its center coincides with the center of the quadrangle*

There exists some degenerative case when $\lambda = 90°$. Then

$$p_2 = p_1 \cos 2\theta + \sin^2\theta$$

So for any $p_1$ and any $\theta$ the value $p_2$ is in [0,1] but has not all possible values from it. The points $p_1, p_2$ are on the direct line:

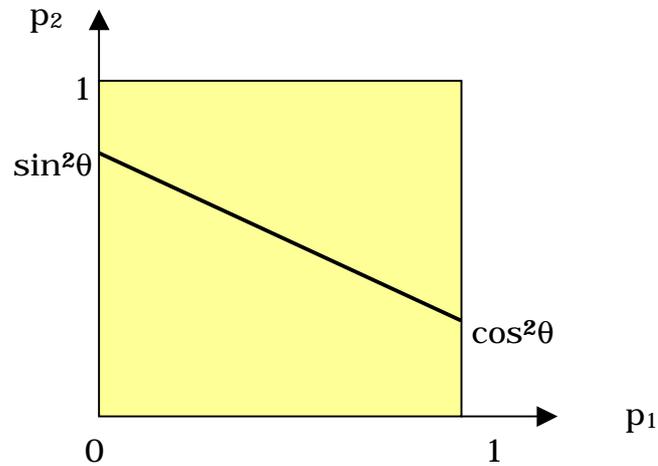

Figure 3. *The degenerative case*